\documentclass[slac_one]{revtex4}
\usepackage{graphicx}
\usepackage{fancyhdr}
\begin{document}

\title{Flavour Violation in GMSB Scenarios: Constraints and Phenomenology} 

\author{Bj\"orn Herrmann}
\affiliation{Laboratoire de Physique Subatomique et de Cosmologie,
UJF -- CNRS\,/\,IN2P3 -- INPG,\\
53 Avenue des Martyrs, 38026 Grenoble, France}

\begin{abstract}
We present an extensive analysis of low-energy, electroweak precision, and cosmological constraints in the Minimal Supersymmetric Standard Model (MSSM) with Gauge-Mediated Supersymmetry Breaking (GMSB) and including the possibility of Non-Minimal Flavour Violation (NMFV). Using detailed scans over the parameter space, we show how NMFV allows to relax the stringent constraints excluding large parts of the minimal GMSB. We define benchmark scenarios and present numerical predictions for various sparticle production cross sections at the LHC.
\end{abstract}

\maketitle

\thispagestyle{fancy}

\section{CONSTRAINTS ON THE MINIMAL GMSB \label{sec2}}

In the absence of experimental evidence for Supersymmetry (SUSY), it is essential to constrain the parameter space of the Minimal Supersymmetric Standard Model (MSSM) both at the SUSY-breaking and the electroweak scale. The underlying SUSY-breaking mechanism plays a key role in the phenomenology of weak-scale SUSY. Gauge-Mediated SUSY Breaking (GMSB) is an attractive scenario, where the breaking is mediated from the secluded to the observable sector through a gauge-singlet chiral superfield and $n_q$ quark-like and $n_l$ lepton-like messenger fields \cite{GMSB,GiudiceRattazzi}. The phenomenology is determined by the effective SUSY-breaking scale $\Lambda$, the messenger mass scale $M_{\rm mes}$, the messenger index $N_{\rm mes}$, the ratio of the vacuum expectation values of the two Higgs doublets, $\tan\beta$, and the sign of the higgsino mass parameter $\mu$. Another free parameter is the gravitino mass $m_{\tilde{G}}$, related to the SUSY-breaking scale $\langle F\rangle$ and the Planck mass $M_{\rm P}$ through $m_{\tilde{G}}=\langle F\rangle/(\sqrt{3} M_{\rm P})$.

A rather stringent constraint on the model comes from the theoretically robust inclusive branching ratio
\begin{equation}
  {\rm BR}(b\to s\gamma) ~=~ \big( 3.55 \pm 0.26 \big) \cdot 10^{-4} ,
\end{equation}
obtained from the combined measurements of BaBar, Belle, and CLEO \cite{HFAG}. A second observable influenced by the SUSY masses and mixings is the electroweak $\rho$-parameter, where new physics contributions are constrained to \cite{PDG2006}
\begin{equation}
  \Delta\rho ~=~ \big( 1.02 \pm 0.86 \big) \cdot 10^{-3} .
\end{equation}
As a third constraint, we require the SUSY contribution to the anomalous magnetic moment of the muon, $a^{\rm SUSY}_{\mu}$, to close the gap between the experimental measurements and Standard Model (SM) prediction \cite{PDG2006}, 
\begin{equation}
  \Delta a_{\mu} ~=~ a_{\mu}^{\rm exp} - a_{\mu}^{\rm SM} ~=~ \big( 29.2 \pm 8.6 \big) \cdot 10^{-10} .
\end{equation}
Since the SUSY one-loop contribution favours positive values of the higgsino mass parameter $\mu$ \cite{Moroi}, we do not consider the case of $\mu<0$ in our analysis. Finally, we require the gravitino to be the candidate for cold dark matter in the Universe, i.e.\ its mass should be larger than $10^{-4}$ GeV and its relic density $\Omega_{\tilde{G}}h^2$ should agree with the current limit
\begin{equation}
  0.094 \le \Omega_{\tilde{G}}h^2 \le 0.136 ,
\end{equation}
obtained  at the $2\sigma$ confidence level from the WMAP mission in combination with SDSS and BAO data \cite{HamannWMAP}. For a given scenario, the calculation of $\Omega_{\tilde{G}}h^2$ involves the gluino and gravitino masses, the relic density of the next-to-lightest SUSY particle (NLSP), and the reheating temperature $T_{\rm R}$ of the Universe \cite{GravitinoRelic}. For the latter, values of $T_{\rm R} \gtrsim 10^9$ GeV are preferred in scenarios with leptogenesis \cite{Leptogenesis}. The last constraint concerns the abundances of the light elements in the Universe, which may be spoiled if the NLSP lives too long before decaying into the gravitino. We therefore require the NLSP lifetime to be shorter than $6\cdot 10^3$ seconds \cite{NLSPLifetime}.

We impose the above electroweak and low-energy constraints at the $2\sigma$ level on the minimal GMSB parameter space. Starting from the reduced set of high-scale parameters we perform the renormalization group running with {\tt SPheno 2.2.3} \cite{SPheno} and compute the mass eigenvalues and the observables ${\rm BR}(b\to s\gamma)$, $\Delta\rho$, and $a_{\mu}^{\rm SUSY}$ with {\tt FeynHiggs 2.6.4} \cite{FeynHiggs}. In the left panel of Fig.\ \ref{fig1}, we show a typical scan of the $\Lambda$--$M_{\rm mes}$ plane for $\tan\beta=15$, $N_{\rm mes}=1$, and $\mu>0$ revealing that ``collider-friendly'' regions with small or intermediate SUSY masses are virtually excluded by the severe constraint from $b\to s\gamma$ \cite{FuksGMSB}. The region $\Lambda > M_{\rm mes}$ is excluded due to unphysical solutions to the RGE.

\section{FLAVOUR VIOLATION IN GMSB MODELS \label{sec3}}

Although the minimal GMSB is known to suppress flavour-changing neutral currents and thus avoid the ``SUSY flavour problem'' arising naturally in gravity-mediated models, models beyond the minimal GMSB can reintroduce flavour-breaking terms at the electroweak scale \cite{GiudiceRattazzi,FVinGMSB,DubovskyGorbunov}. For our study, we consider the possibility of mixing between messenger and matter fields \cite{DubovskyGorbunov}, introducing flavour violation either only in the left-left or both in the left-left and right-right chiral squark sectors depending on the nature of the messengers. We here focus on fundamental messengers inducing flavour mixing for left-handed squarks only. The complete analysis including antisymmetric messengers leading to flavour violation in both the left-left and right-right chiral sectors can be found in Ref.\ \cite{FuksGMSB}. Note that we do not consider slepton flavour violation.

The resulting flavour-violating terms have to be included in the squark mass matrices at the electroweak scale. Since flavour mixing is less constrained between the second and third generations \cite{NMFV23gen} we do not consider mixing with first generation squarks. Non-minimal flavour violation is then implemented through one dimensionless real parameter $\lambda_{\rm LL}$, that parametrizes the off-diagonal entries $\Delta^{23}_{\rm LL}$ of the squared squark mass matrices in terms of the corresponding soft SUSY-breaking diagonal elements, $\Delta^{23}_{\rm LL} = \lambda_{\rm LL} M_{\tilde{q}_2} M_{\tilde{q}_3}$. The diagonalization of these matrices leads to the physical squark mass eigenstates labeled $\tilde{q}_1,\dots,\tilde{q}_6$ for $q=u,d$ ordered by increasing masses. For a more detailed discussion of the implementation of non-minimal flavour violation in the MSSM see e.g.\ Refs.\ \cite{BozziMSUGRA,FuksGMSB,YellowReport}.

\begin{figure}
  \includegraphics[scale=0.25]{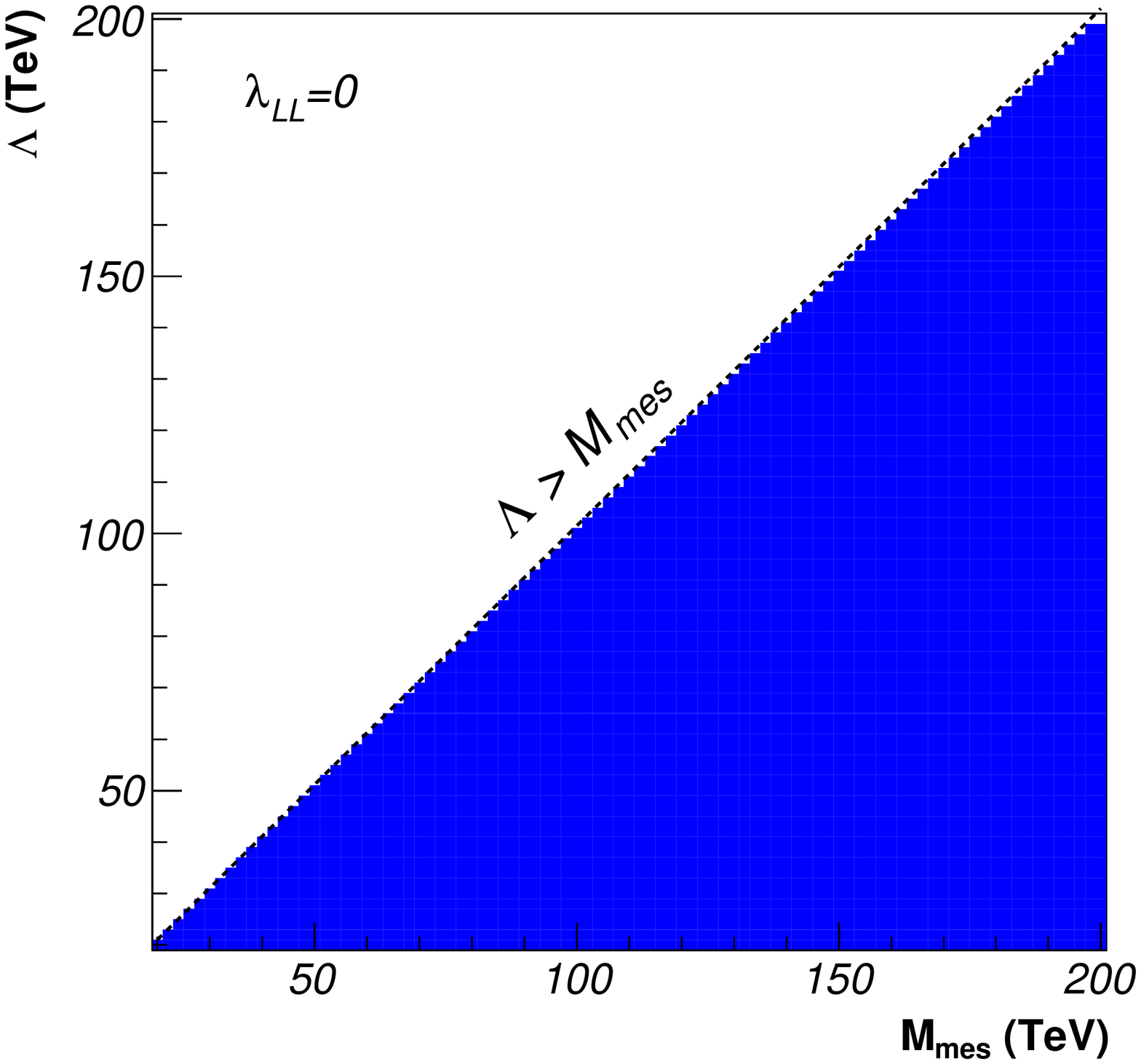}
  \includegraphics[scale=0.25]{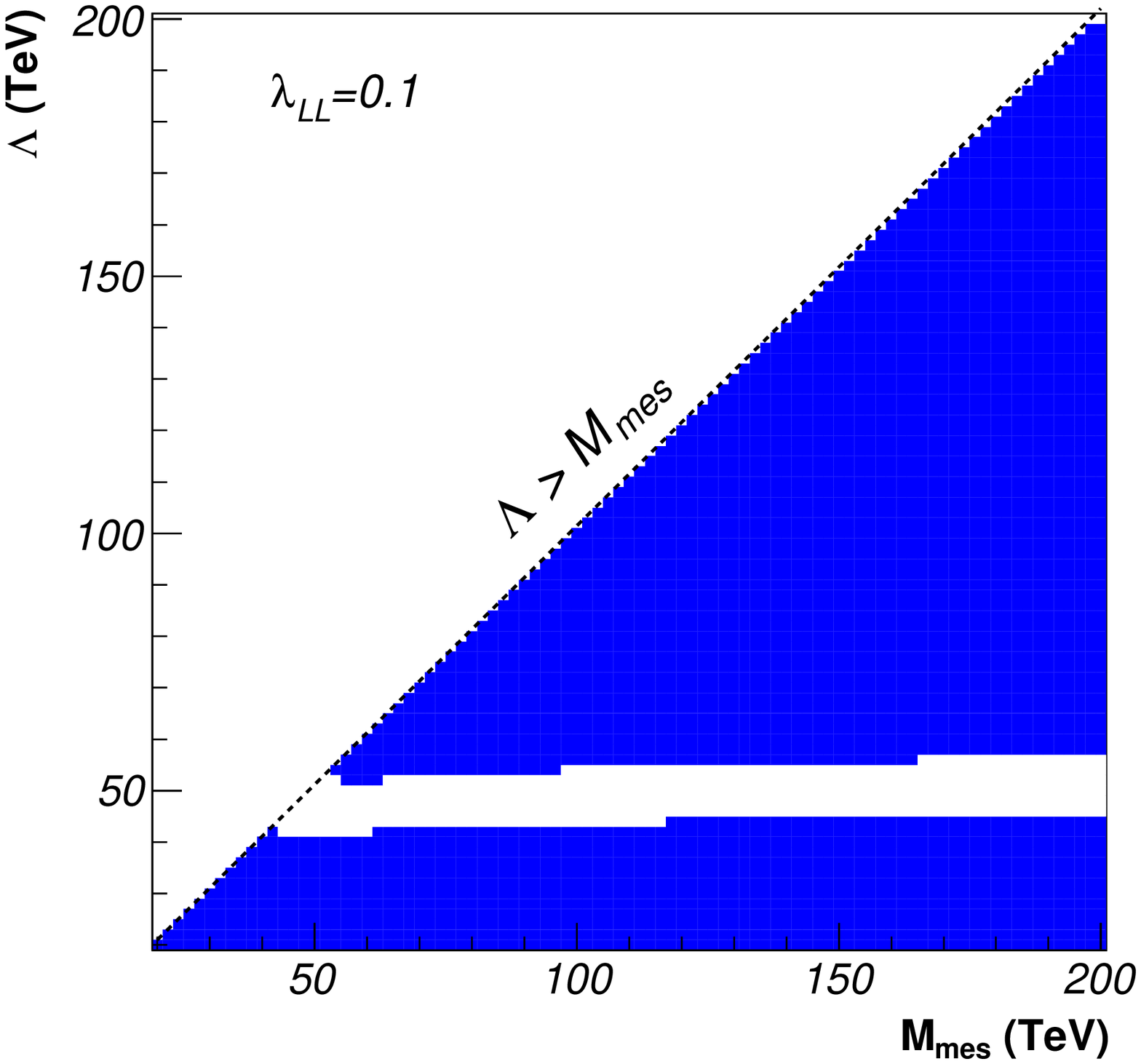}
  \includegraphics[scale=0.25]{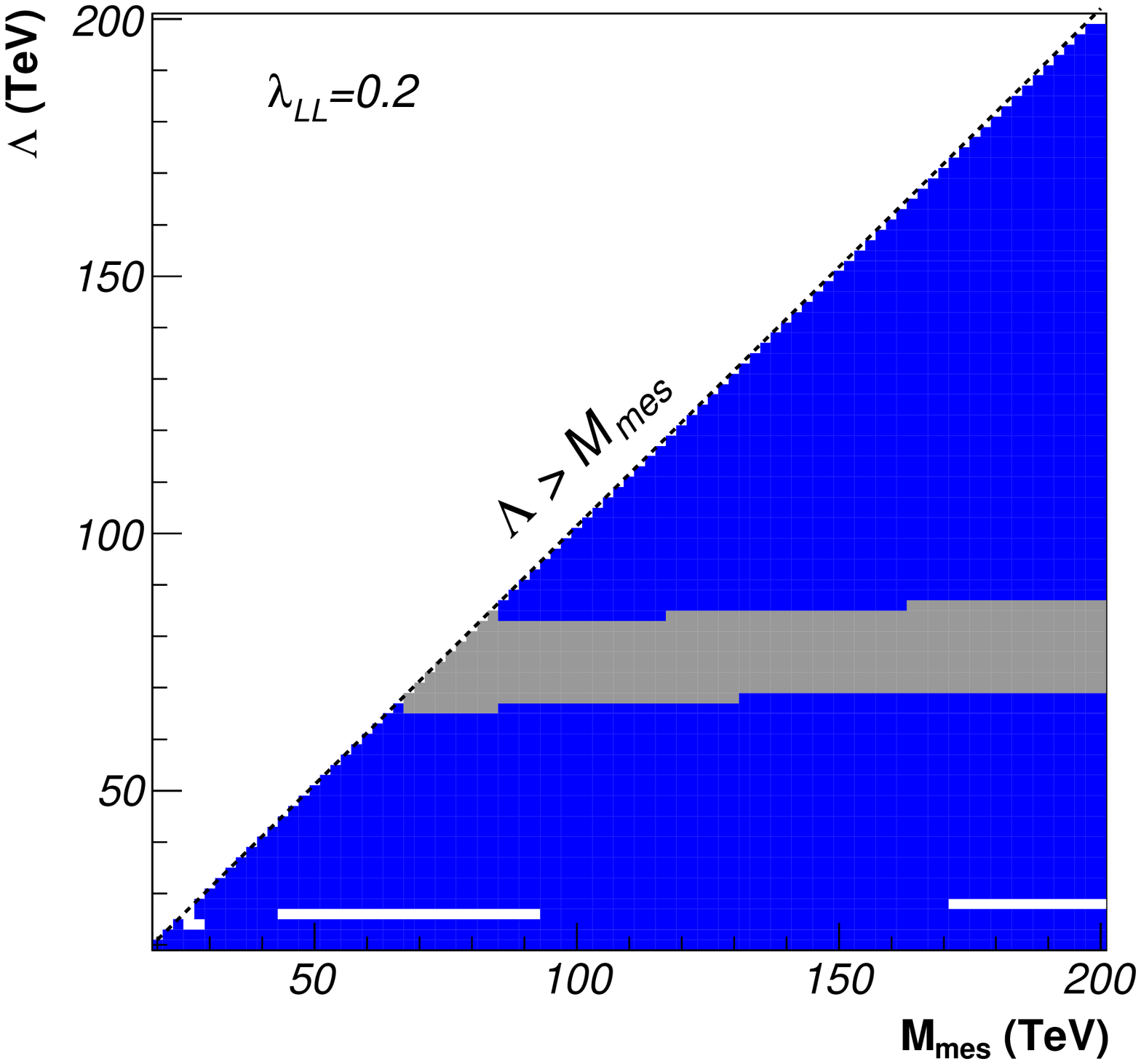}
  \vspace*{-3mm}
  \caption{The regions excluded by $b\to s\gamma$ (dark) and favoured by $a_{\mu}$ (grey) in the $\Lambda$--$M_{\rm mes}$ plane of the minimal ($\lambda_{\rm LL}=0$) and non-minimal flavour violating ($\lambda_{\rm LL}>0$) GMSB parameter space at $\tan\beta=15$, $N_{\rm mes}=1$, $\mu>0$. \label{fig1}}
\end{figure}

We now rescan the $\Lambda$--$M_{\rm mes}$ plane for values of $\lambda_{\rm LL} \lesssim 0.2$. The resulting excluded or favoured regions with respect to $b\to s\gamma$ and $a_{\mu}$ are shown in the central and right panels of Fig.\ \ref{fig1} for $\lambda_{\rm LL}=0.1$ and $0.2$, respectively. The figures show that flavour mixing between second and third generation squarks opens windows in the ``collider-friendly'' regions of the GMSB parameter space that are both allowed by $b\to s\gamma$ and favoured by $a_{\mu}$. In this region, we define a benchmark point at $\Lambda=65$ TeV, $M_{\rm mes}=90$ TeV, $N_{\rm mes}=1$, $\tan\beta=15$, and $\mu>0$. Further possibilities involving other values of $\tan\beta$ and $N_{\rm mes}$ are discussed in Ref.\ \cite{FuksGMSB}.

\begin{figure}
  \includegraphics[scale=0.25]{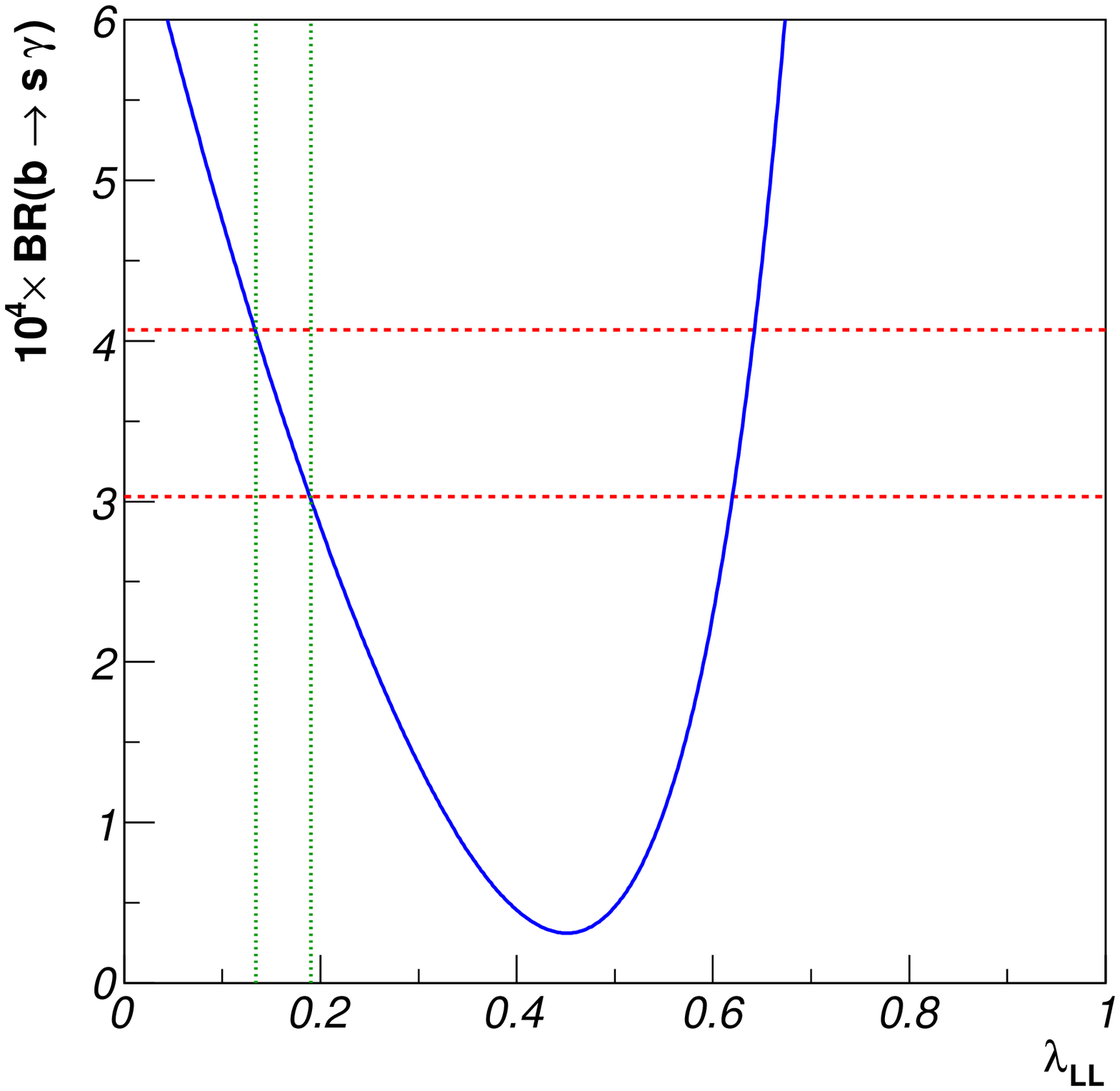}
  \includegraphics[scale=0.25]{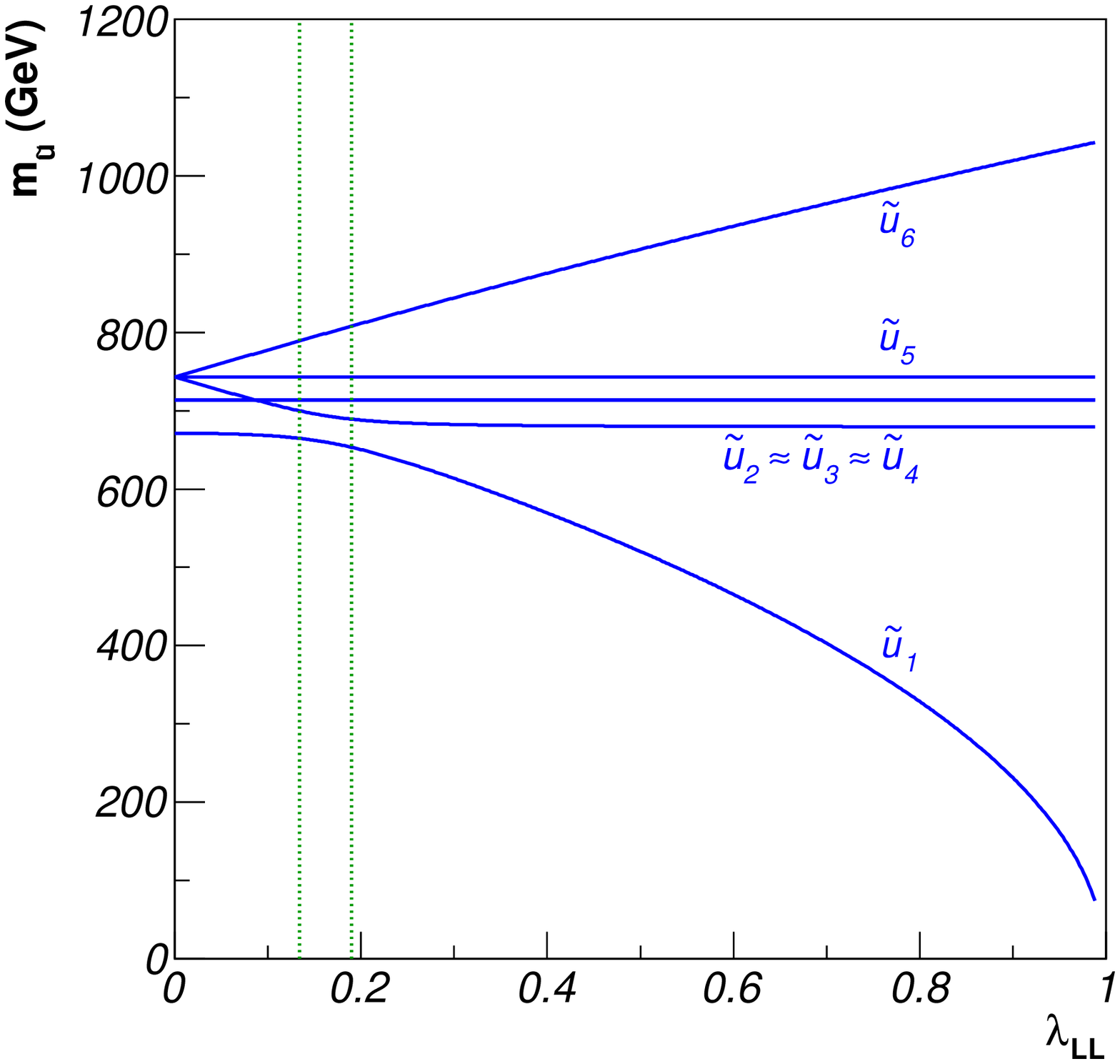}
  \includegraphics[scale=0.25]{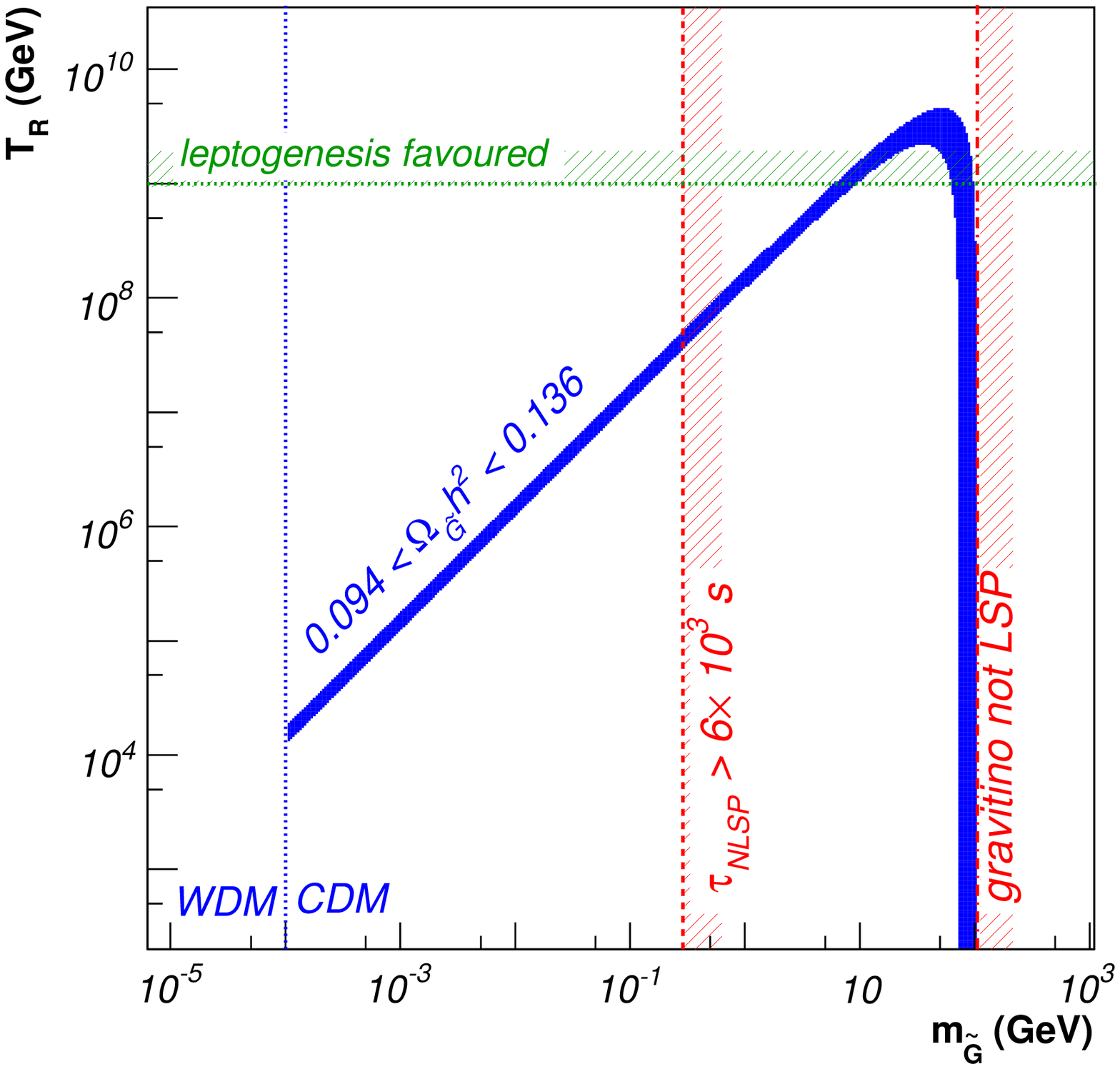}
  \vspace*{-3mm}
  \caption{The branching ratio ${\rm BR}(b\to s\gamma)$ (left) and the mass eigenvalues of up-type squarks (centre) as function of the flavour-violation parameter $\lambda_{\rm LL}$ and the cosmological constraints on the $m_{\tilde{G}}$--$T_{\rm R}$ plane for the discussed scenario. \label{fig2}}
\end{figure}

For this benchmark scenario, we now inspect in detail the experimental constraints and the squark mass eigenvalues as a function of the flavour-violation parameter $\lambda_{\rm LL}$. Since the squark two-loop contributions are suppressed with respect to the one-loop slepton diagrams, the anomalous magnetic moment is $a^{\rm SUSY}_{\mu} = 37.7 \cdot 10^{-10}$ independently of $\lambda_{\rm LL}$. Contrary, the observable $\Delta\rho$ depends strongly on squark mass splitting and thus on flavour violation. The large experimental uncertainties, however, allow for rather high values of $\lambda_{\rm LL} \lesssim 0.5$. The most stringent constraint comes from the decay $b\to s\gamma$, for which we show the branching ratio in the first panel of Fig.\ \ref{fig2}. The small experimentally allowed range, indicated at $2\sigma$ by two horizontal dotted lines, allows for two narrow intervals for $\lambda_{\rm LL}$. Since the one around $\lambda_{\rm LL} \approx 0.62$ is disfavoured by $b\to s\mu^+\mu^-$ measurements \cite{GambinoBSmumu}, we obtain the allowed range $0.14 \le \lambda_{\rm LL} \le 0.2$ for the flavour-violation parameter, indicated by two vertical lines. 

Let us now turn to the up-type squark mass eigenvalues shown in the central panel of Fig.\ \ref{fig2}. For higher values of $\lambda_{\rm LL}$, we observe an important splitting between the lightest and the heaviest squark due to the increasing off-diagonal elements of the mass matrix. An interesting phenomenon of level-reordering between neighbouring states can be observed at lower values of $\lambda_{\rm LL}$. These ``avoided crossings'' are a common behaviour for Hermitian matrices that depend continuously on a single real parameter. At the point where two levels should cross the corresponding squarks exchange their flavour content, which leads to an interesting phenomenology in the context of squark production discussed in Sec.\ \ref{sec4}. The same phenomena are observed for down-type squarks, that are not shown here.

In the last panel of Fig.\ \ref{fig2} we show the cosmologically favoured or excluded regions of the $m_{\tilde{G}}$--$T_{\rm R}$ plane according to the constraints discussed in Sec.\ \ref{sec2}. We cannot fulfill all the constraints at a time since the constraints on the relic density and NLSP lifetime are in conflict with high reheating temperatures $T_{\rm R} \gtrsim 10^9$ GeV. We therefore allow for lower temperatures and fix the gravitino mass to be of the order of $10^{-1}$ GeV, which satisfies the constraints for still relatively high values of $T_{\rm R} \sim 10^7$ GeV.

\section{PREDICTIONS FOR THE LHC \label{sec4}}

We finally present numerical predictions for the production cross sections of squarks and gauginos at the LHC, i.e.\ for $pp$-collisions at $\sqrt{s}=14$ TeV centre-of-momentum energy. Unpolarized hadronic cross sections are obtained by convolving the relevant partonic cross section with the universal parton density functions (PDF) according to the QCD factorization theorem. We here employ the leading-order (LO) set of the CTEQ6 global parton density fit \cite{CTEQ6}. Analytical expressions for partonic cross sections including NMFV can be found in Refs.\ \cite{BozziMSUGRA,FuksGMSB}.

\begin{figure}
  \includegraphics[scale=0.25]{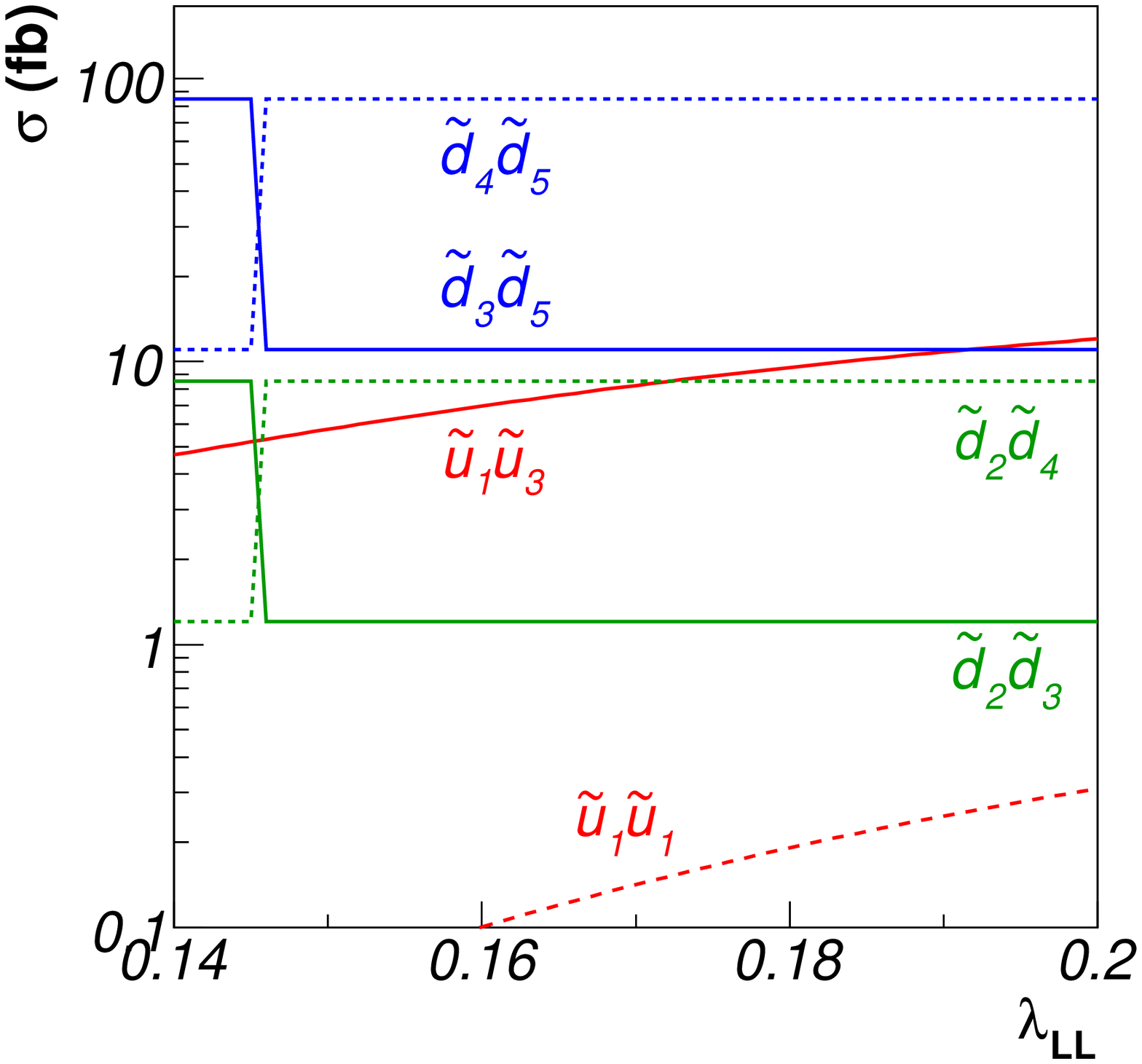}
  \includegraphics[scale=0.25]{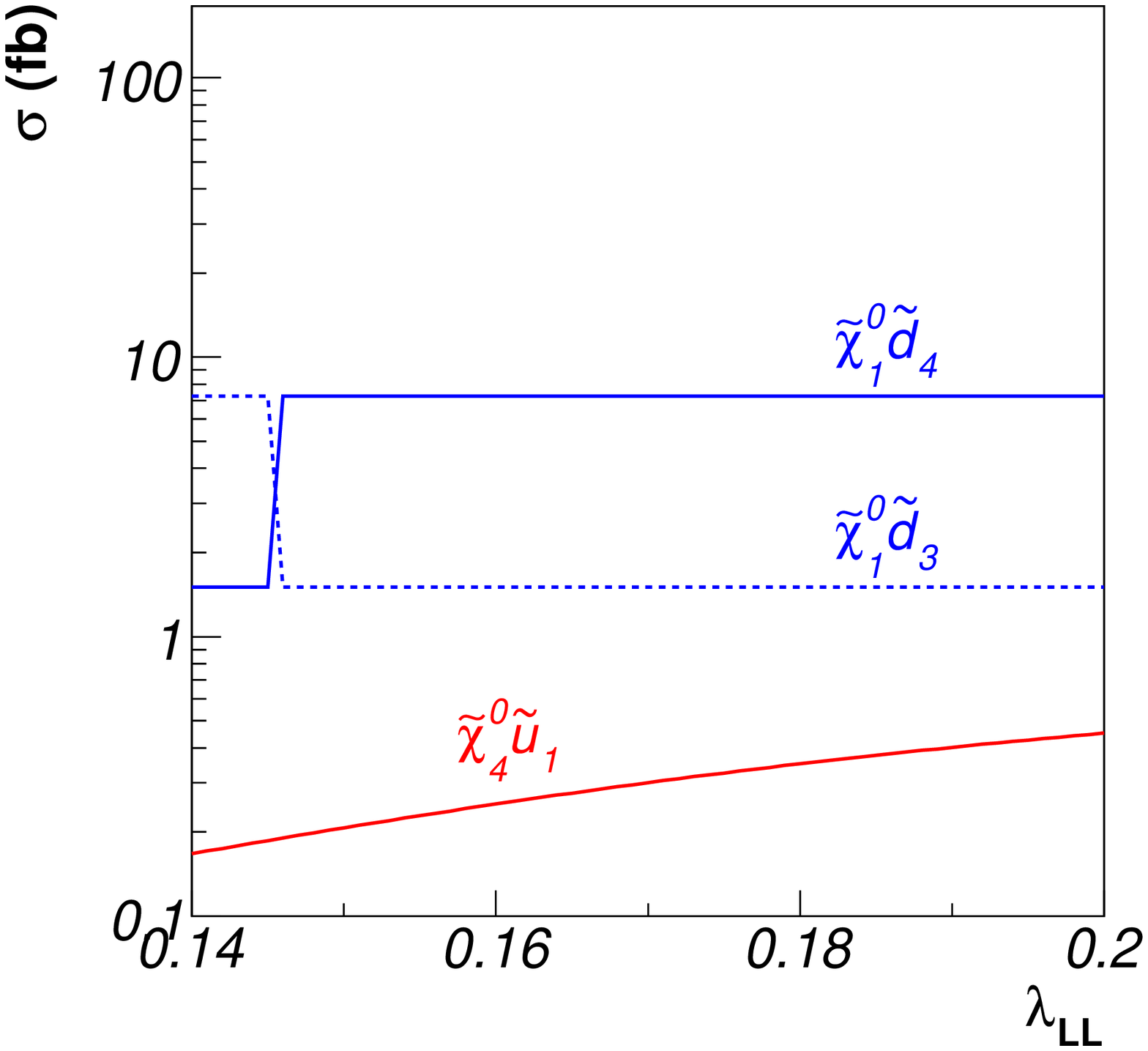}
  \includegraphics[scale=0.25]{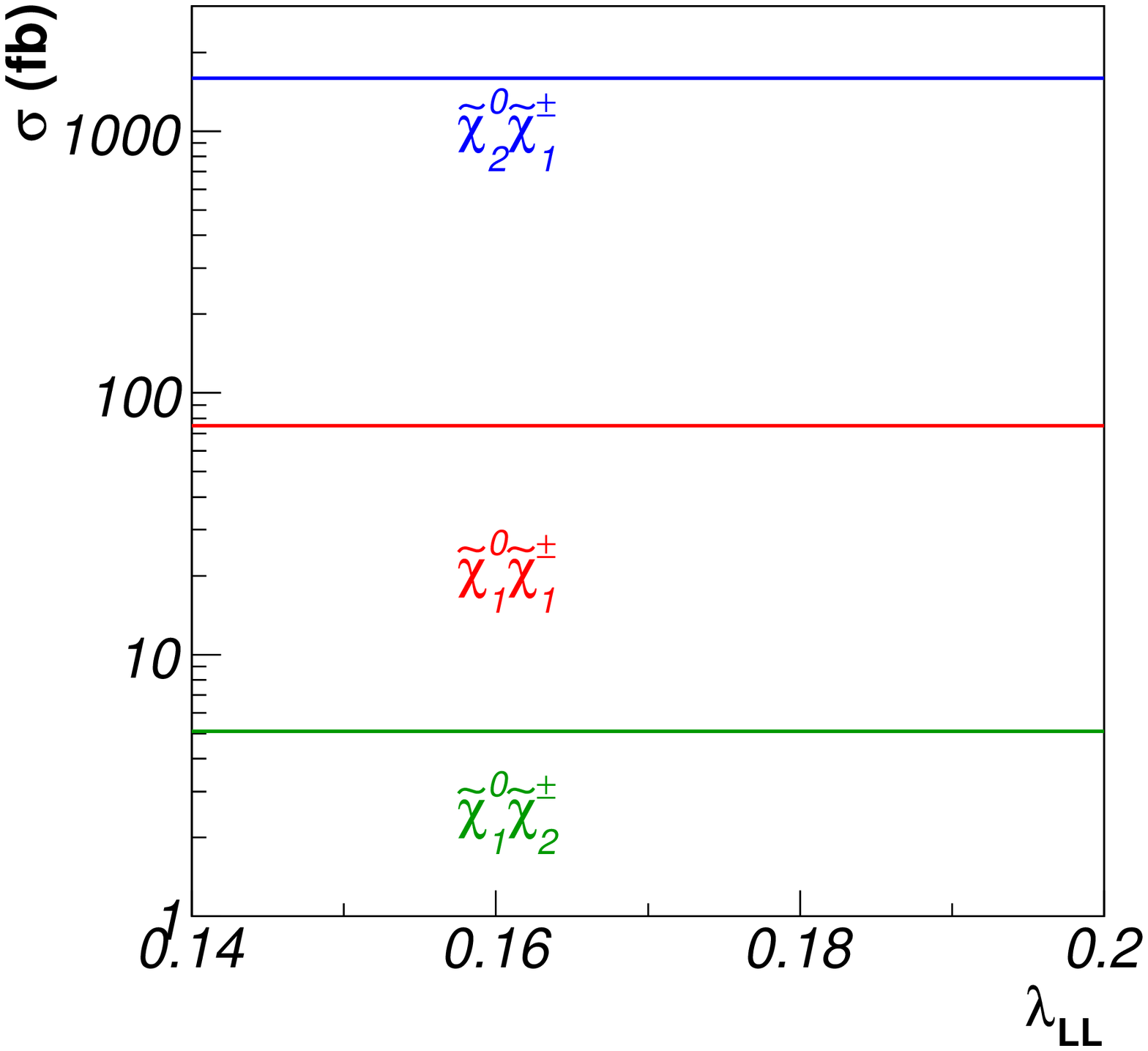}
  \vspace*{-4mm}
  \caption{Examples for squark pair, squark-gaugino associated, and gaugino pair production cross sections at the LHC. \label{fig3}}
\end{figure}

In Fig.\ \ref{fig3}, we show examples for squark pair production (left), squark-gaugino associated production (centre), and gaugino-pair production (right) for our benchmark scenario with flavour violation in the left-left chiral squark sector. A more complete compilation of cross sections including other benchmark scenarios can be found in Ref.\ \cite{FuksGMSB}. In the left panel of Fig.\ \ref{fig3}, we observe cross sections that increase with our flavour-violation parameter $\lambda_{\rm LL}$. This behaviour is explained by changes of the squark flavour content with increasing flavour mixing. For lower values of $\lambda_{\rm LL}$ the squark $\tilde{u}_1$ is mostly stop-like, so that its production would necessite a top quark in the initial state. However, for larger values of $\lambda_{\rm LL}$, $\tilde{u}_1$ receives a sizeable scharm content and, due to the non-zero PDF of the charm quark in the proton, the cross sections of e.g.\ the squark pairs $\tilde{u}_1\tilde{u}_1$ and $\tilde{u}_1\tilde{u}_3$ increase.

Other production cross sections show sharp transitions in particular production channels at a precise value of the flavour-violation parameter, as can be seen e.g.\ for $\tilde{d}_3\tilde{d}_5$ and $\tilde{d}_4\tilde{d}_5$ at $\lambda_{\rm LL} \approx 0.145$. The two cross sections exchange their values, which is due to an "avoided crossing" occuring here and inducing the exchange of the flavour contents of the two squarks. For $\lambda_{\rm LL} < 0.145$, $\tilde{d}_3$ is purely sdown-like and the large down-quark density in the proton leads to important cross sections. At $\lambda_{\rm LL} \approx 0.145$, however, the flavour content of $\tilde{d}_3$ changes and it becomes sstrange-like. The corresponding cross sections are then lower due to the smaller PDF of strange quarks. The squark $\tilde{d}_4$ receives the inverse flavour change from sstrange-like for $\lambda \lesssim 0.145$ to sdown-like for $\lambda_{\rm LL} \gtrsim 0.145$. In consequence, its production cross sections increase dramatically at this value of the flavour violation parameter.

The same phenomena are observed for the associated production of squarks and gauginos shown in the central panel of Fig.\ \ref{fig3}. In the right panel of Fig.\ \ref{fig3} we show the production of gaugino pairs, where flavour violation is involved through squark-quark-gaugino vertices. However, squarks only appear as propagators in $t$- or $u$-channel diagrams, and the resulting sum over all mass eigenstates renders the cross section independent of $\lambda_{\rm LL}$.

Concerning the production of gravitinos, which is potentially interesting, unfortunately our scenarios featuring a rather heavy gravitino ($m_{\tilde{G}}\sim 10^{-1}$ GeV) do not allow for observable cross sections at the LHC. Only a very light gravitino ($m_{\tilde{G}} \lesssim 1$ eV) would lead to sizeable production cross sections. 



\begin{thebibliography}{99} 

\bibitem{GMSB}
M.~Dine, A.~E.~Nelson, 
Phys. Rev. D {\bf 48} 1277, 1993; 
M.~Dine, A.~E.~Nelson, A.~Shirman, 
Phys. Rev. D {\bf 51} 1362, 1995;
M.~Dine, A.~E.~Nelson, Y.~Nir, A.~Shirman, 
Phys. Rev. D {\bf 53} 2658, 1996.

\bibitem{GiudiceRattazzi}
G.~F.~Giudice, R.~Rattazzi, 
Phys. Rept. {\bf 322} 419, 1999.

\bibitem{HFAG}
E.~Barbiero {\it et al.} (Heavy Flavour Averaging Group (HFAG)), 
hep-ex/0603003.

\bibitem{PDG2006}
W.~M.~Yao {\it et al.} (Particle Data Group), 
J. Phys. G {\bf 33} 1, 2006, and 2007 partial update.

\bibitem{Moroi}
T.~Moroi, 
Phys. Rev. D {\bf 53} 6565, 1996.

\bibitem{HamannWMAP}
J.~Hamann, S.~Hannestad, M.~S.~Sloth, Y.~Y.~Y.~Wong, 
Phys. Rev. D {\bf 75} 023522, 2007.

\bibitem{GravitinoRelic}
M.~Bolz, A.~Brandenburg, W.~Buchm{\"u}ller, Nucl. Phys. B {\bf 606} 518, 2001,
[Erratum-ibid. B {\bf 790} 336, 2008];
J.~Pradler, F.~Steffen, Phys. Rev. D {\bf 75} 023509, 2007;
V.~S.~Rychkov, A.~Strumia, Phys. Rev. D {\bf 75} 075011, 2007.

\bibitem{Leptogenesis}
W.~Buchm{\"u}ller, P.~Di Bari, M.~Pl{\"u}macher,
Annals Phys. {\bf 315} 305, 2005.

\bibitem{NLSPLifetime}
M.~Pospelov, J.~Pradler and F.~D.~Steffen,
arXiv:0807.4287 [hep-ph].

\bibitem{SPheno}
W.~Porod, 
Comp. Phys. Comm. {\bf 153} 275, 2003.

\bibitem{FeynHiggs}
S.~Heinemeyer, W.~Hollik, G.~Weiglein, 
Comp. Phys. Comm. {\bf 124} 76, 2000.

\bibitem{FuksGMSB}
B.~Fuks, B.~Herrmann, and M.~Klasen, 
arXiv:0808.1104 [hep-ph].

\bibitem{FVinGMSB}
K.~Tobe, J.~D.~Wells, T.~Yanagida, 
Phys. Rev. D {\bf 69} 035010, 2004.

\bibitem{DubovskyGorbunov}
S.~L.~Dubovsky, D.~S.~Gorbunov, 
Nucl. Phys. B {\bf 557} 199, 1999.

\bibitem{NMFV23gen}
F.~Gabbiani, E.~Gabrielli, A.~Masiero, L.~Silvestrini, 
Nucl. Phys. B {\bf 477} 321, 1996;
T.~Hahn, W.~Hollik, J.~I.~Illana, S.~Penaranda, 
hep-ph/0512315;
J.~Foster, K.~I.~Okumura, L.~Roszkowski, 
Phys. Lett B {\bf 641} 452, 2006;
M.~Chiuchini, A.~Masiero, P.~Paradisi, L.~Silvestrini, S.~K.~Vempati, O.~Vives, 
Nucl. Phys. B {\bf 783} 112, 2007.

\bibitem{BozziMSUGRA}
G.~Bozzi, B.~Fuks, B.~Herrmann, and M.~Klasen, 
Nucl. Phys. B {\bf 787} 1, 2007.

\bibitem{YellowReport}
F.~del Aguila {\it et al.},
arXiv:0801.1800 [hep-ph].

\bibitem{GambinoBSmumu}
P.~Gambino, U.~Haisch, M.~Misiak,
Phys. Rev. Lett. {\bf 94} 061803, 2005.

\bibitem{CTEQ6}
J.~Pumplin, D.~R.~Stump, J.~Huston, H.~L.~Lai, P.~Nadolsky, W.~K.~Tung,
JHEP {\bf 0207} 012, 2002.


\end{thebibliography}
\end{document}